\documentstyle[12pt]{article}

\advance \textheight by \topskip
\makeatletter
\@addtoreset{equation}{section}

\makeatother

\begin{document}

\title{Classical Anomalies for Spinning Particles}
\author{Jorge Gamboa${}^1${}\thanks{e-mail: jgamboa@lauca.usach.cl}
$$ 
and Mikhail Plyushchay${}^{2,3}${}\thanks{e-mail: plyushchay@mx.ihep.su}\\
\smallskip\\
{\it ${}^1$ Departamento de F\'{\i}sica, Universidad de Santiago de
Chile}\\ 
{\it Casilla 307, Santiago 2, Chile}\\
{\it ${}^2$Institute for High Energy Physics, Protvino}\\ 
{\it Moscow Region, 142284 Russia}\\
{\it ${}^3$Departamento de Fisica,
Universidade Federal de Juiz de Fora}\\
{\it 36036-330 Juiz de Fora, MG Brazil}}

\date{}

\maketitle

\begin{abstract}
We discuss the phenomenon of classical anomaly.  It is observed for
$3D$ Berezin-Marinov (BM), Barducci-Casalbuoni-Lusanna (BCL) and
Cort\'es-Plyushchay-Vel\'azquez (CPV) pseudoclassical spin particle
models. We show that quantum mechanically these different models
correspond to the same $P,T$-invariant system of planar fermions, but
the quantum system has global symmetries being not reproducible
classically in full in any of the models.  We demonstrate that
the specific U(1) gauge symmetry characterized by the opposite
coupling constants of spin $s=+1/2$ and $s=-1/2$ states has a natural
classical analog in CPV model but can be reproduced in BM and BCL
models in an obscure and rather artificial form. We also show that BM
and BCL models quantum mechanically are equivalent in any
odd-dimensional space-time, but describe different quantum systems in
even space-time dimensions.
\vskip 0.2cm
\noindent
PACS numbers: 12.90.+b, 11.30.-j
\vskip 0.2cm
\noindent 
Keywords: spin particle models, (super)symmetries, anomalies.
\vskip 0.2cm
\begin{center}
{\bf Nucl. Phys. B 512 (1998) 485-504}
\end{center}
\end{abstract}
\newpage
\section{Introduction}

When global or local symmetries do not survive under quantization,
i.e.  some classical symmetries are lost at the quantum level, we
have global \cite{wi} or local quantum anomalies \cite{loc,ht}.  Since the
quantization is not unique, different quantum systems corresponding
to one and the same classical system can be constructed and sometimes
symmetries can be restored by modifying appropriately the
quantization procedure.  {\it A priori} there is no obstruction for
existence of the inverse picture when different classical models
correspond to one and the same quantum system having symmetries to be
not reproducible at least in some of these classical models.

The present paper is devoted to investigation of such a phenomenon of
{\it classical anomaly} for the $P,T-$invariant $3D$ fermion system 
\cite{highT,ps,gps}.

The phenomenon of classical anomaly will be revealed here in the
following aspect.  We shall investigate classical and quantum
theory of three $3D$ spin particle models.  These are the
Berezin-Marinov (BM)
\cite{bm,betal,gt}, Barducci-Casalbuoni-Lusanna (BCL)
\cite{bcl} and Cort\'es-Plyushchay-Vel\'azquez (CPV)
\cite{cpv,gps} pseudoclassical models.  All these three models have
the same number of gauge symmetries generated by the corresponding
first class constraints.  Dirac quantization procedure supplies us in
all three cases with the mass shell condition and with the
$P,T$-invariant system of $3D$ Dirac equations which are the quantum
counterparts of the corresponding classical constraints.  Therefore,
the quantization of these different models leads to one and the same
quantum system of the $P,T-$invariant massive planar fermions.  We
shall get the set of integrals of motion of the quantum system
(generators of global symmetries) and establish the (super)algebras
formed by them.  Then we shall find that

1) not all the quantum integrals of motion have classical analogs
in CPV model,

2) though BCL and BM models contain all the formal classical analogs
of the quantum integrals, nevertheless the corresponding symmetry
(super)algebras can be reproduced classically in BCL model only
partially, whereas the BM model does not allow us to reproduce even a
part of corresponding (super)algebras.  On the other hand, we shall
see that some properties of (super)symmetry generators being
not reproducible in BCL model, are reproduced classically in CPV
model.

Therefore, we shall show that all the three models reflect
classically some different parts of properties of the
corresponding quantum system and neither of them reproduce the
quantum symmetry properties in full.

In addition, we shall demonstrate that the so called $\sigma_3$ (or
$\tau_3$ in terminology of Refs. \cite{highT}) local U(1) gauge
symmetry, characterized by the opposite coupling constants of spin
$s=+1/2$ and $s=-1/2$ states and denoted further as
U${}_{\sigma_3}$(1), has a natural classical analog in CPV model but
can be reproduced in BM and BCL models in an obscure and rather
artificial form.

Just from the described features of classical anomaly taking
place for the concrete system to be investigated here it is
clear how important role this phenomenon may play.
Indeed, if the quantization of some model gives the necessary
quantum system, it may turn out that not all the quantum
symmetry properties of the system are automatically
reflected at the classical level. On the other hand, it may turn
out that in order to reflect classically all the quantum
symmetry properties of the system, it is necessary to use
different classical models taking into account different
aspects of symmetry properties of the same quantum system.  We shall
discuss other  possible consequences of the observed phenomenon in
last Section.

The paper is organized as follows. In Section 2 we analyze the
classical properties of BM and BCL models.  We show that these models
are closely related but not equivalent: the BCL model has one odd
integral of motion which is absent from the BM model. This difference
has no consequences from the point of view of quantum theory of the
models in the case of odd-dimensional space-time but turns out to be
crucial under quantization in even space-time dimensions, in
particular, in $3+1$ dimensions where the models were constructed
originally.  The quantization of $3D$ BM and BCL models
by the Dirac method is realized in Section 3.  In Section 4 we
consider the CPV model and compare it with the BM and BCL
models. The peculiarity of the CPV model is that it does not
admit, even local, gauge conditions \cite{pr}.  In Section 5 we discuss
the quantum symmetries of the corresponding $P,T$-invariant planar
fermion system.  Section 6 is devoted to the discussion of classical
anomalies.  In Section 7 we analyze the classical counterparts of the
U${}_{\sigma_3}$(1) gauge theory.  Section 8 contains some concluding
remarks and discussion. Appendix is devoted to the quantization
of BM and BCL models in even-dimensional space-time.

\section{BM and BCL models}
\subsection{Lagrangians}
The Lagrangians of BM and BCL models, introduced originally 
in $3+1$ dimensions, 
in general case of $D$-dimensional space-time are given by
\begin{equation}
L_{BM}=\frac{1}{2e}(\dot{x}_\mu+i\lambda\xi_\mu)^2
-\frac{e}{2}m^2-im\lambda\xi_*-\frac{i}{2}\xi_\mu\dot{\xi}{}^\mu
-\frac{i}{2}\xi_*\dot{\xi}_*,
\label{bm0}
\end{equation}
\begin{equation}
L_{BCL}=\frac{1}{2e}(\dot{x}_\mu+im^{-1}\dot{\xi}_*\xi_\mu)^2
-\frac{e}{2}m^2-\frac{i}{2}\xi_\mu\dot{\xi}{}^\mu
+\frac{i}{2}\xi_*\dot{\xi}_*,
\label{bcl0}
\end{equation}
where $x_\mu$ are even coordinates of the particle of mass $m$,
$\xi_\mu$ and $\xi_*$ are Grassmann (odd) spin variables,
$e$ and $\lambda$ are even and odd Lagrange multipliers,
respectively, and we use the metric
$\eta_{\mu\nu}=diag(-,+,...,+)$.
Under space inversion the variables $e,$ $\lambda$ and 
$\xi_*$ 
are transformed as a scalar, $P: e\rightarrow e$, and
as pseudoscalars,
$P:\xi_*\rightarrow-\xi_*$,
$P: \lambda\rightarrow -\lambda$,
whereas $x_\mu$ and $\xi_\mu$ are treated as 
vector and pseudovector quantities.
In particular cases of 2+1  and 3+1 dimensions
this correspondingly means that
$
P: x_\mu\rightarrow (x_0,-x_1,x_2)$,
$P: \xi_\mu\rightarrow -(\xi_0,-\xi_1,\xi_2)$
and 
$
P: x_\mu\rightarrow (x_0,-x_1,-x_2,-x_3)$,
$P: \xi_\mu\rightarrow -(\xi_0,-\xi_1,-\xi_2,-\xi_3)$.

These models are usually referred to as one and the same model.
Indeed, the Lagrangians can be equivalently represented as
\begin{equation}
L_{BM}=L_0-i(e^{-1}\dot{x}\xi-m\xi_*)\lambda,
\label{bm1}
\end{equation}
\begin{equation}
L_{BCL}=L_0-i(e^{-1}\dot{x}\xi-m\xi_*)m^{-1}\dot{\xi}_*,
\label{bcl1}
\end{equation}
where 
\begin{equation}
L_0=\frac{1}{2e}\dot{x}{}^2-\frac{e}{2}m^2
-\frac{i}{2}\xi_\mu\dot{\xi}{}^\mu-\frac{i}{2}\xi_*\dot{\xi}_*.
\label{l0}
\end{equation}
Therefore, from here one could conclude that after
identification 
\begin{equation}
\lambda=m^{-1}\dot{\xi}_* 
\label{laxi}
\end{equation}
we have equivalent models.
However, we shall see that this is not so: the non-equivalence of the
models reveals itself in different and specific ways at the classical
and quantum levels.   Classically this becomes clear if we note that
the BCL Lagrangian (\ref{bcl0}) unlike the Lagrangian of BM model
(\ref{bm0}) is quasi-invariant with respect to the transformations
\begin{equation}
\xi_*\rightarrow\xi_*+\rho \Rightarrow
L_{BCL}\rightarrow L_{BCL}+
\frac{d}{d\tau}\left(\frac{i}{2}\rho\xi_*\right),
\label{quasi}
\end{equation}
where $\rho$ is an arbitrary odd real constant parameter.
BCL Lagrangian, unlike that of BM model, 
is also quasi-invariant under the
global transformations 
\begin{equation}
\delta x_\mu=-im^{-1}\xi_*\epsilon_\mu,\quad
\delta\xi_\mu=\epsilon_\mu,\quad \delta\xi_*=0
\label{susy?}
\end{equation}
with odd real constant vector $\epsilon_\mu$.
The first Noether's theorem \cite{Dirac}
gives us the associated scalar and vector 
integrals of motion being the generators
of the transformations (\ref{quasi}) and (\ref{susy?}),
\begin{equation}
\theta_0=\frac{1}{2}\xi_*+i\pi_*,\quad
\Sigma_\mu=m\xi_\mu+\xi_* p_\mu,
\label{suin}
\end{equation}
where $\pi_*=\partial L_{BCL}/\partial \dot{\xi}_*$ and
$p_\mu=\partial L_{BCL}/\partial \dot{x}{}^\mu$. 
The commutator of two supertranslations specified by the parameters
$\rho_1,$ $\epsilon^\mu_1$ and $\rho_2$, $\epsilon_2^\mu$
gives
\begin{equation}
\delta_{1,2}
x^\mu=im^{-1}(\epsilon_1^\mu \rho_2-\epsilon_2^\mu \rho_1),\quad
\delta_{1,2}\xi_\mu=\delta_{1,2}\xi_*=0,
\label{?su}
\end{equation}
where
$\delta_{1,2}=\delta_2\delta_1-\delta_1\delta_2$.
Therefore, the commutator of supertransformations (\ref{quasi}),
(\ref{susy?}) produces the space-time translation. This could be
considered as an indication on some global space-time supersymmetry
in BCL model.  We shall discuss this point in detail in what follows.

\subsection{Relationship of the models: Lagrangian formalism}
Let us investigate the difference between BM and BCL systems in
more detail.  First, we shall show that classically BCL model
has effectively one more odd dynamical variable in comparison
with BM model, and this additional variable is a constant of
motion generating the transformations (\ref{quasi}).  
As a result, quantum theory of both models will not be
equivalent in the case of even dimension of space-time, whereas
in odd space-time dimension it will be the same.  In spite of
the quantum equivalence in latter case, we shall observe that BM
model cannot reproduce classically some quantum symmetries being
reproducible in BCL model.

Lagrange equations of motion of BCL model are
\begin{eqnarray}
&\dot{p}_\mu=0,\quad p^2+m^2=0,&
\label{e1}\\
&\dot{\xi}_\mu+m^{-1}\dot{\xi}_*p_\mu=0,&
\label{e1+}\\
&\frac{d}{d\tau}(\xi_*-m^{-1}p\xi)=0,&
\label{e2}
\end{eqnarray}
where 
\begin{equation}
p_\mu=e^{-1}(\dot{x}_\mu+im^{-1}\dot{\xi}_*\xi_\mu).
\label{p1}
\end{equation}
Equations of motion for BM model have the same form (\ref{e1})
but with $p_\mu$ given by 
\begin{equation}
p_\mu=e^{-1}(\dot{x}_\mu+i\lambda\xi_\mu),
\label{p2}
\end{equation}
whereas instead of Eqs. (\ref{e1+}), (\ref{e2}) we have 
\begin{eqnarray}
&\dot{\xi}_\mu+\lambda e^{-1}\dot{x}_\mu=0,&
\label{e2+*}\\
&
\dot{\xi}_*-m\lambda=0,&
\label{e2*}
\end{eqnarray}
and variation in $\lambda$ gives Lagrange constraint
\begin{equation}
\xi_*-m^{-1}p\xi=0.
\label{con1}
\end{equation}
Solving Eq. (\ref{e2*}) with respect to 
odd Lagrange multiplier, we arrive exactly at 
the relation (\ref{laxi})
and conclude that energy-momentum vectors given by
Eqs. (\ref{p1}) and (\ref{p2}) are identical,
and as a consequence, Eqs. (\ref{e1+})
are identical to Eqs. (\ref{e2+*}), 
whereas Eq. (\ref{e2}) is a derivative 
of Eq. (\ref{con1}).
This means that for BM model we have the relation
\begin{equation}
p\xi-m\xi_*=0,
\label{?1}
\end{equation}
whereas for BCL model,
\begin{equation}
p\xi-m(\xi_*+\rho)=0,
\label{?2}
\end{equation}
where $\rho$ is an arbitrary real odd constant in
correspondence with symmetry (\ref{quasi}) observed above.

Formally the classical non-equivalence happens here since for 
excluding Lagrange multiplier $\lambda$
from BM model we have to use the equation of motion for
$\xi_*$ but not the equation for $\lambda$ itself
(see the discussion of this point in Refs. \cite{gt,ht}).

\subsection{Hamiltonian formalism} 

Let us consider the Hamiltonian description of the systems.
The BM model is described by the canonical even variables
$x_\mu$, $p^\mu$, $\{x_\mu,p^\nu\}=\eta_\mu^\nu$,
by $e$, $P_e$, $\{e,P_e\}=1$,
and by odd canonical variables $\xi_\mu$ and
$\pi_\mu$, $\{\xi_\mu,\pi_\nu\}=-\eta_{\mu\nu}$,
$\xi_*$ and  $\pi_*$, $\{\xi_*,\pi_*\}=-1$,
$\lambda$ and $\Pi_\lambda$,
$\{\lambda,\Pi_\lambda\}=-1$.
First-class primary constraints are
\begin{equation}
P_e\approx 0,\quad
\Pi_\lambda\approx 0,
\label{pi1}
\end{equation}
and second-class primary constraints are
\begin{equation}
\pi_*-\frac{i}{2}\xi_*\approx 0,
\label{se1}
\end{equation}
\begin{equation}
\pi_\mu-\frac{i}{2}\xi_\mu\approx 0.
\label{se2}
\end{equation}
Taking into account second-class constraints
removes odd canonical momenta $\pi_\mu$, $\pi_*$, and
gives the Dirac brackets
\begin{equation}
\{\xi_\mu,\xi_\nu\}=i\eta_{\mu\nu},
\label{d1}
\end{equation}
\begin{equation}
\{\xi_*,\xi_*\}=i.
\label{d2}
\end{equation}
Stability algorithm for the constraints
(\ref{pi1}) generates secondary constraints
\begin{equation}
\phi=\frac{1}{2}(p^2+m^2)\approx 0,
\quad
\chi=p\xi-m\xi_*\approx 0.
\label{constr}
\end{equation}
The set of constraints (\ref{pi1}),
(\ref{constr}) is the set of first class constraints
with constraints (\ref{constr}) forming
$s(1)$ superalgebra \cite{CrRit}:
$\{\chi,\chi\}=2i\phi$,
$\{\chi,\phi\}=0$. 
The total Hamiltonian \cite{ht,Dirac} is 
\begin{equation}
H_{tot}=e\phi-i\lambda\chi+wP_e+\omega\Pi_\lambda,
\label{?3}
\end{equation}
where $w=w(\tau)$ and $\omega=\omega(\tau)$ are real
even and odd arbitrary functions, respectively.
Equations of motion generated by Hamiltonian (\ref{?3}),
\begin{eqnarray}
&\dot{p}_\mu=0,\quad
\dot{x}_\mu=ep_\mu-i\lambda\xi_\mu,\quad \dot{e}=w,&
\label{?4}\\
&\dot{\xi}_*=m\lambda,\quad
\dot{\xi}_\mu=-\lambda p_\mu,\quad
\dot{\lambda}=\omega,&
\label{?5}
\end{eqnarray}
are equivalent to Lagrange equations of motion.

In the case of BCL model the primary constraints are
$P_e\approx 0$,
\begin{equation}
{\chi}=p\xi-m\left(\frac{1}{2}\xi_*-i\pi_*\right)\approx 0,
\label{tch}
\end{equation}
and those given by Eq. (\ref{se2}). 
Taking into account second class constraints
(\ref{se2}) excludes $\pi_\mu$ and gives again Dirac brackets 
(\ref{d1}). 
The condition $\dot{P}_e=0$ generates
the mass shell constraint as a secondary constraint,
and taking into account corresponding brackets,
we find that constraints $\phi$ and ${\chi}$
form the same $s(1)$ superalgebra as $\phi$ and $\chi$ do
in BM model.
The total Hamiltonian
\begin{equation}
H_{tot}=e\phi+wP_e-i\tilde{\omega}{\chi}
\label{thbcl}
\end{equation}
containing an arbitrary odd function 
$\tilde{\omega}=\tilde{\omega}(\tau)$
generates the equations of motion
equivalent to Lagrange equations of motion.
Let us define the real odd variables
\begin{equation}
\theta_0=\frac{1}{2}\xi_*+i\pi_*,\quad
\theta_1=\frac{1}{2}\xi_*-i\pi_*
\label{th}
\end{equation}
equivalent to the initial set of scalar odd variables 
$\xi_*$ and $\pi_*$. 
They have nontrivial Poisson brackets
\begin{equation}
\{\theta_\alpha,\theta_\beta\}=i\eta_{\alpha\beta},
\label{thbr}
\end{equation}
where $\alpha,\beta=0,1$, $\eta_{\alpha\beta}=diag(-,+)$.
Therefore, odd constraint has the form 
\begin{equation}
{\chi}=p\xi-m\theta_1\approx 0.
\label{chth}
\end{equation}
Taking into account the form of the corresponding 
brackets and constraints as well as total Hamiltonians,
one concludes that the Lagrange multiplier $\lambda$
of BM model can be identified with arbitrary odd function
$\tilde{\omega}$ from BCL model,
whereas the odd scalar space-like variable $\theta_1$ of the
BCL model corresponds to the scalar variable 
$\xi_*$ of the BM model.
Then the difference of the BCL model from the BM model consists in
the presence of the additional time-like odd variable $\theta_0$
being a constant of motion which generates the transformation
(\ref{quasi}).

\subsection{Integrals of motion}

In addition to $\theta_0$,
other classical integrals of motion of BCL model are
the energy-momentum vector $p_\mu$,
the total angular momentum tensor
\[
J_{\mu\nu}=x_\mu p_\nu-x_\nu p_\mu+i\xi_\mu\xi_\nu,
\]
and odd vector
\begin{equation}
\Xi_\mu=m\xi_\mu+\theta_1 p_\mu.
\label{xi}
\end{equation}
However, not all the components of $\Xi_\mu$ are independent since
taking into account the mass shell constraint, we have
$p^\mu\Xi_\mu\approx m\chi\approx 0$.  
Due to the constraint $\chi\approx 0$, we
have also the integral of motion $i\theta_1p\xi$ which classically
(but not quantum mechanically) is weakly equal to zero.  In the case
of BM model all the listed integrals except $\theta_0$ (with
corresponding change of $\theta_1$ for $\xi_*$) are the same.

As we stressed above, the integral $\theta_0$ generates in BCL model
the symmetry transformations (\ref{quasi}).
On the other hand, the vector integral $\Sigma_\mu$
being the generator of supersymmetry transformations
(\ref{susy?}), is presented as
\begin{equation}
\Sigma_\mu=\Xi_\mu +\theta_0 p_\mu.
\label{Sig}
\end{equation}
We see that the generator $\Sigma_\mu$ is the linear combination of
the integral $\Xi_\mu$ and of the composition of odd, $\theta_0$, and
even, $p_\mu$, integrals. Therefore, the integral $\Sigma_\mu$ plays
no special independent role and the nature of the global
supersymmetry (\ref{?su}) turns out to be trivial: it is encoded in
the relation $\{\Sigma_\mu,\theta_0\}=-ip_\mu$ being a consequence of
Eq. (\ref{Sig}).  Here we note that symmetry (\ref{susy?}) presents
also in BM model under taking into account the equation of
motion (\ref{e2*}). Its generator is the odd vector
integral $\Xi_\mu=m\xi_\mu+\xi_* p_\mu.$ Thus, we arrive once more at
the conclusion that the essential difference of pseudoclassical BM
model from the BCL model is due to the absence in it of the odd
integral $\theta_0$.

To conclude the discussion of classical theory of BM and BCL
models, we recall that according to the general theory of
constrained dynamical systems
\cite{ht,Dirac} any phase space function $A$ such that
$\frac{d}{d\tau}A=0$
is the integral of motion, where $\tau$ is an evolution parameter.
Any integral of motion can be considered as a generator of
corresponding symmetry transformation.  Observables are those phase
space functions satisfying the relations $\{A,\phi_a\}\approx 0$, where
$\phi_a$ is the set of constraints.  In our case the Hamiltonians are
linear combinations of constraints and, therefore, the observables are
simultaneously integrals of motion.

\section{Quantization}
\subsection{Quantum equivalence of $3D$ BM and BCL models} 
Let us consider the quantization of both models,
and we start from the BM model.
First we note that the
pair of conjugate odd variables
$\lambda$ and $\Pi_\lambda$ can be completely 
removed from the theory by introducing 
the gauge condition 
$\lambda-\lambda_0\approx 0$
to the constraint $\Pi_\lambda\approx 0$,
where $\lambda_0$ is some odd constant.
Then, constructing quantum analogs of odd variables
$\xi_\mu$ and $\xi_*$ we should consider separately the cases
of odd and even dimensions of space-time. We discuss here the most
interesting case of $3$-dimensional space-time.  Other
odd-dimensional cases can be considered in analogous way, whereas the
case of even space-time dimension will be analyzed in Appendix.

The quantization of odd variables should give us
irreducible representation of the Clifford algebra
with 4 generators,
\[
[\widehat{\xi}_\mu,\widehat{\xi}_\nu]_+=-\eta_{\mu\nu},\quad
[\widehat{\xi}_*,\widehat{\xi}_*]_+=-1,\quad
[\widehat{\xi}_\mu,\widehat{\xi}_*]_+=0,
\]
which is 4-dimensional
(in general case of $d=2n-1$, the dimensionality of the
corresponding Clifford algebra with $2n$ generators is $2^n$).
It is convenient to realize $\widehat{\xi}_\mu$ and $\widehat{\xi}_*$
as 
\begin{equation}
\widehat{\xi}_\mu=\frac{1}{\sqrt{2}}\gamma_\mu\otimes\sigma_1,\quad
\widehat{\xi}_*=\frac{i}{\sqrt{2}}1\otimes\sigma_2,
\label{?7}
\end{equation}
where $(2+1)$-dimensional matrices
$\gamma^\mu$, $\gamma^0=\sigma_3$,
$\gamma^j=i\sigma_j$, $j=1,2$, satisfy the relation
$
\gamma^\mu\gamma^\nu=-\eta^{\mu\nu}+i\epsilon^{\mu\nu\lambda}
\gamma_\lambda
$
with totally antisymmetric tensor $\epsilon^{\mu\nu\lambda}$ 
normalized as $\epsilon^{012}=1.$
The quantum analog of the odd constraint function is
$\frac{1}{\sqrt{2}}1\otimes\sigma_1 \cdot 
(p\gamma\otimes 1+m\cdot 1\otimes\sigma_3),$ and therefore,
the quantum constraint equation $\widehat{\chi}\Psi=0$ is equivalent to
the
pair of Dirac equations,
\begin{equation}
(p\gamma\otimes 1+m\cdot 1\otimes\sigma_3)\Psi=0,
\label{D3}
\end{equation}
where we suppose that $\Psi$ is a doublet of spinor fields,
presented in transposed form as
$\Psi=(\psi_u,\psi_d)$,
$1\otimes\sigma_3\cdot \Psi=(\psi_u,-\psi_d)$.
Here and in what follows we denote the operators $p_\mu$ 
and $x_\mu$
in the same way as their classical counterparts.
Eq. (\ref{D3}) generates the Klein-Gordon equation
being the quantum analog of the constraint $\phi\approx 0$,
and we conclude that the quantization
of BM model in odd dimensional space-time gives the
$P,T$-invariant system of two fermion fields.

Let us consider the quantum theory corresponding to the BCL model
in odd-dimensional space-time.
In this model we have the odd variable
$\theta_0$ in addition to the variables $\xi_\mu$
and $\xi_*$ from BM model. The latter, as it was noted,
should be identified with $\theta_1$.
The quantum analog
of additional odd variable can be realized as
\begin{equation}
\widehat{\theta}_0=\frac{1}{\sqrt{2}}1\otimes\sigma_3,
\label{?8}
\end{equation}
whereas all other variables can be realized exactly
as in BM model.
Therefore, in 2+1 dimensions the quantum analog of the odd
constraint gives here, again, the $P,T$-invariant pair of Dirac
equations.  Since in BCL model we have the Clifford algebra with
odd number of generators, in $2+1$ dimensions we have a relation
being specific to its irreducible representations:
\begin{equation}
2^{5/2}i\widehat{\theta}_0\widehat{\theta}_1\widehat{\xi}{}^0
\widehat{\xi}{}^1\widehat{\xi}{}^2=1.
\label{quant1}
\end{equation}
Therefore, using it, at the quantum mechanical level
we can `restore' the operator $\widehat{\theta}_0$
being absent from the BM model:
\begin{equation}
\widehat{\theta}_0=2^{3/2}i\widehat{\xi}_*\widehat{\xi}{}^0
\widehat{\xi}{}^1\widehat{\xi}{}^2.
\label{t0}
\end{equation}
Because of this, {\it quantum mechanically}
BM and BCL models are equivalent in $(2+1)$-dimensional space-time
as well as in any other odd space-time dimension.
But since there is no classical analog of relation
(\ref{quant1}), classically these models cannot be considered
as equivalent. 

We conclude that BM and BCL models in 2+1 dimensions describe the
same quantum $P,T$-invariant system of two fermion fields.

\subsection{Scalar product}

Under quantization the scalar product should be introduced
in such a way that all the quantum counterparts of classical
real observables would be self-conjugate operators.
In the case of BM model
the choice 
$\langle\Psi_1\vert\Psi_2\rangle=
\bar\Psi_1\Psi_2$ with $\bar\Psi=\Psi^\dagger\Delta$,
$\Delta=\sqrt{2}\widehat{\xi}{}^0$,
guarantees that the quantum operators corresponding
to the observables $J_{\mu\nu}$ and $\Xi_\mu$
will be self-conjugate:
$\langle\Psi_1\vert \widehat{O}\Psi_2\rangle^*=
\langle\Psi_2\vert \widehat{O}{}^{\star}\Psi_1\rangle
=\langle\Psi_2\vert \widehat{O}\Psi_1\rangle$,
$\widehat{O}=\widehat{J}_{\mu\nu},\widehat{\Xi}_\mu$.
We find also that with such a choice the averaged
odd quantum constraint gives the standard form for the
field Lagrangian of the $3D$ $P,T$-invariant fermion system:
\begin{equation}
\langle\Psi\vert\widehat{\chi}\Psi\rangle=
\Psi^\dagger\gamma^0(p\gamma\otimes 1+m1\otimes\sigma_3)
\Psi={\cal L}.
\label{flan}
\end{equation}
Note that the operator $i\widehat{\theta}_0$
constructed according to the relation (\ref{t0})
is self-conjugate here but having no classical analog.

In the case of BCL model we find that the same indefinite scalar
product with $\Delta=\sqrt{2}\widehat{\xi}{}^0$, guarantees that
operators $\widehat{J}_{\mu\nu}$ and $\widehat{\Xi}_\mu$ are
self-conjugate
and that it produces the same field Lagrangian under averaging the quantum
constraint $\widehat{\chi}$.  But in this case we get that
$\widehat{\theta}_0$, being observable, is an anti-self-conjugate
operator.  In the BCL model we cannot introduce the scalar product in
such a way that all the quantum analogs of classical real observable
quantities would be self-conjugate operators, and in this sense the
model reveals some sort of quantum anomaly. 

\section{CPV model}
\subsection{Classical theory}
The third pseudoclassical model to be considered here 
is the $3D$ CPV model \cite{cpv} given by the Lagrangian
\begin{equation}
L_{CPV}=\frac{1}{2e}(\dot{x}_\mu+iv\epsilon_{\mu\nu\lambda}\xi^\nu
\xi^\lambda)^2
-\frac{e}{2}m^2-imv\theta_1\theta_2
-\frac{i}{2}\xi_\mu\dot{\xi}{}^\mu
+\frac{i}{2}\theta_a\dot{\theta}_a
\label{?10}
\end{equation}
with $v$ being even Lagrange multiplier instead of odd
multiplier $\lambda$ taking place in BM model,
and $\theta_a$, $a=1,2,$ being the pair of 
odd Lorentz scalars. We assume that under $P$-inversions here,
unlike the BM and BCL models, 
$\xi_\mu$ is transformed as vector $x_\mu$, 
$\theta_1$ is a scalar variable,
whereas $\theta_2$ and $v$ are pseudoscalar quantities
\cite{gps}.  
Due to the presence of even Lagrange multiplier instead
of odd one the correspondence between
CPV and BM (as well as BCL) models has a formal character 
given by relations 
$\lambda\xi_\mu
\leftrightarrow 
v\epsilon_{\mu\nu\lambda}
\xi^\nu\xi^\lambda$,
$\lambda\xi_* 
\leftrightarrow
v\theta_1\theta_2$,
$\xi_*\dot{\xi}_*
\leftrightarrow
-\theta_a\dot{\theta}_a$.
The formal character of the relationship 
between models is
clear also from the fact that CPV model,
unlike BM and BCL models,
has a global U(1) invariance,
\begin{equation}
\theta^\pm\rightarrow
\theta'=e^{\pm i\omega}\theta^\pm,
\label{u1}
\end{equation}
where $\theta^\pm=\frac{1}{\sqrt{2}}(\theta_1\pm i\theta_2)$.  As we
shall see, this global invariance can be naturally localized in CPV
model leading to the quantum U${}_{\sigma_3}$(1) gauge symmetry
\cite{highT,gps}.

Dirac brackets for $\xi_\mu$ have the same form (\ref{d1})
as in BM and BCL models, whereas for $\theta_a$ we have
\begin{equation}
\{\theta_a,\theta_b\}=-i\delta_{ab}.
\label{?11}
\end{equation}
The primary Hamiltonian constraints are $P_e\approx0$, $P_v\approx0$
with $P_v$ being canonical momentum for $v$,
$\{v,P_v\}=1$. As a secondary constraint we get the mass shell constraint
$\phi=\frac{1}{2}(p^2+m^2)\approx 0$,
and instead of odd constraint $\chi$ from BM and BCL models, we have
here nonlinear in Grassmann variables even constraint
\begin{equation}
\varrho=i\epsilon_{\mu\nu\lambda}
p^\mu\xi^\nu\xi^\lambda-2mi\theta_1\theta_2\approx0.
\label{nonl}
\end{equation}
The Hamiltonian is
\[
H=e\phi+v\varrho+u_1P_e+u_2P_v,
\]
where $u_{1,2}=u_{1,2}(\tau)$ are arbitrary even functions.
Due to nonlinearity of constraint (\ref{nonl})
in Grassmann variables, it admits no gauge conditions.
The general class of the systems with such peculiar
constraints was investigated in Ref. \cite{pr}.

\subsection{Integrals of motion}
The vectors
$p_\mu$ and 
$J_\mu=-\frac{1}{2}\epsilon_{\mu\nu\lambda}J^{\nu\lambda}$,
the latter having the same form as in two other models,
are the integrals of motion.
In addition, we have the following nilpotent
integrals of motion:
$\xi^{(0)}$ and 
\begin{equation}
i\theta_1\theta_2,\quad i\xi^{(1)}\xi^{(2)},\quad
i\xi^{(i)}\theta^i,\quad 
i\epsilon_{ij}\xi^{(i)}\theta^j,
\label{ev}
\end{equation} 
where $\theta^i=\theta_i$, $i,j=1,2$,
$\xi^{(\alpha)}=\xi^\mu e^{(\alpha)}_\mu$, $\alpha=0,1,2$,
$\epsilon_{ij}=-\epsilon_{ji}$, $\epsilon_{12}=1$.
Here having in mind the mass shell constraint,
we introduced the complete oriented triad
$e^{(\alpha)}_{\mu}=e^{(\alpha)}_\mu(p)$, $\alpha=0,1,2$,
\begin{equation}
e^{(0)}_\mu=\frac{p_\mu}{\sqrt{-p^2}},\quad
e^{(\alpha)}_\mu\eta_{\alpha\beta}e^{(\beta)}_\nu=\eta_{\mu\nu},\quad
\epsilon_{\mu\nu\lambda}e^{(0)\mu}e^{(i)\nu}e^{(j)\lambda}=
\epsilon^{0ij}.
\label{tri}
\end{equation}
Note that unlike $e^{(0)}_\mu$,
the components $e^{(i)}_\mu$
are non-covariant objects and as a result,
the same is true for the quantities $\xi^{(i)}$.

In the case of 2+1 dimensions in BM and BCL models
we have two independent components of odd vector integral (\ref{xi}) and
one even nilpotent integral 
$i\xi^{(0)}\theta_1\approx 0$, which are 
supplemented by
one odd scalar integral $\theta_0$
in BCL model. In contrast, the CPV model 
has only one odd scalar integral
$\xi^{(0)}$ and the set of four even nilpotent integrals
(\ref{ev}), two of which on mass shell are related by
the nilpotent constraint (\ref{nonl}): 
$i\xi^{(1)}\xi^{(2)}-i\theta_1\theta_2\approx 0$.

\subsection{Quantization}
The quantum counterparts of odd variables
satisfying the anticommutation relations
$
[\widehat{\theta}_a,\widehat{\theta}_b]_{{}_+}=\delta_{ab},$
$[\widehat{\xi}_\mu,\widehat{\xi}_\nu]_{{}_+}=-\eta_{\mu\nu}$
and
$
[\widehat{\xi}_\mu,\widehat{\theta}_a]_{{}_+}=0$
can be realized as
\[
\widehat{\xi}{}^\mu=\frac{1}{\sqrt{2}}\gamma^\mu\otimes
\sigma_3,\qquad
\widehat{\theta}_a=\frac{1}{\sqrt{2}}1\otimes\sigma_a.
\]
Nilpotent constraint (\ref{nonl}) turns into the equation
\begin{equation}
(p\gamma\otimes 1+m\cdot 1\otimes\sigma_3)\Psi=0,
\label{3dir}
\end{equation}
which itself produces the mass shell (Klein-Gordon) equation.
The scalar product 
\begin{equation}
\langle\Psi_1\vert\Psi_2\rangle=\Psi_1^\dagger \gamma^0\otimes 1\Psi_2
\label{scal0}
\end{equation}
gives the relation of the form (\ref{flan}),
$\frac{1}{2}\langle\Psi\vert\widehat{\varrho}\Psi\rangle={\cal
L}$, and we conclude that {\it quantum mechanically} CPV and
$3D$ BM and BCL models are equivalent: all of them result in the
same $3D$ $P,T$-invariant fermion system.  However it is
necessary to note that unlike the BCL model, all the odd
operators of CPV model, $\widehat{\xi}_\mu$ and
$\widehat{\theta}_a$, are self-conjugate.  This difference is
coded in different nature of classical counterparts of operators
$\widehat{\theta}_a$ and $\widehat{\theta}_\alpha$: variables
$\theta_a$, $a=1,2$, from CPV model are characterized by the
definite metric $\delta_{ab}$ in the sense of brackets
(\ref{?11}), whereas brackets (\ref{thbr}) for variables
$\theta_{\alpha}$, $\alpha=0,1$, from BCL model contain the
indefinite metric $\eta_{\alpha\beta}$.  As we shall see in
Section 6, as a consequence of this formal difference in
properties of odd operators, not all the classical analogs of
corresponding quantum symmetry generators will be real
quantities in $3D$ BCL and BM models.  On the other hand, even
character of classical counterpart (\ref{nonl}) of quantum
constraint (\ref{3dir}) will be the obstacle  for
reproducing classical analog of a part of quantum symmetries
in CPV model.
 
\section{Quantum symmetries}

Let us describe the global symmetries of the  quantum $3D$
$P,T$-invariant fermion system given by Eq. (\ref{3dir}), by the
corresponding field Lagrangian $ {\cal L}=\Psi^\dagger\gamma^0\otimes
1(p\gamma\otimes 1+ m\cdot 1\otimes\sigma_3)\Psi $ and associated
Dirac scalar product (\ref{scal0}).  Eq. (\ref{3dir}) is equivalent
to the Klein-Gordon equation and to the equation $\Pi_+\Psi=0$ with
the operator
$
\Pi_+=\frac{1}{2}(1+\gamma^{(0)}\otimes\sigma_3)
$
being the projector operator, 
$\Pi_+^2=\Pi_+$.
The complete set of observable operators
is the set of operators being self-conjugate 
with respect to the scalar product
(\ref{scal0}) and
commuting with Klein-Gordon operator $p^2+m^2$
and with the operator $\Pi_+$. 
This is the set of operators $p_\mu$, $\widehat{J}_\mu=-
\epsilon_{\mu\nu\lambda}x^\nu p^\lambda-
\frac{1}{2}\gamma_\mu\otimes 1$, 
and 
\begin{equation}
\widehat{S}=-\frac{1}{2}\gamma^{(0)}\otimes 1,\quad
\widehat{N}=\frac{1}{2}1\otimes\sigma_3,\quad
\widehat{T}{}^i=-\frac{1}{2}\gamma^{(i)}\otimes\sigma_1,\quad
\widehat{K}{}^i=-\frac{1}{2}\gamma^{(i)}\otimes\sigma_2.
\label{ob0}
\end{equation}
Though operators $\widehat{T}{}^i$ and $\widehat{K}{}^i$
are related by unitary transformation, nevertheless 
the usage of both sets of operators will be necessary in what
follows. 
Operators $\widehat{\cal W}{}^{\alpha}$, $\alpha=0,1,2$,
where $\widehat{\cal W}{}^{0}=\widehat{S}$,
and $\widehat{\cal W}{}^i=\widehat{T}{}^i$,
or $\widehat{\cal W}{}^{i}=\widehat{K}{}^i$,
and operators $\widehat{\cal N}{}^\alpha$, where 
$\widehat{\cal N}{}^0=-\widehat{N}$ and $\widehat{\cal
N}{}^1=\widehat{T}{}^1$, 
$\widehat{\cal N}{}^2=\widehat{K}{}^1$, or
$\widehat{\cal N}{}^1=\widehat{T}{}^2$, $\widehat{\cal
N}{}^2=\widehat{K}{}^2$,
form $su(1,1)$ algebras, 
\[
[\widehat{\cal W}_\alpha,\widehat{\cal
W}_\beta]=-i\epsilon_{\alpha\beta\gamma}
\widehat{\cal W}{}^{\gamma},\quad
[\widehat{\cal N}_\alpha,\widehat{\cal N}_\beta]=-i
\epsilon_{\alpha\beta\gamma}\widehat{\cal N}{}^\gamma.
\]
Simultaneously, they are generators of
Clifford algebras $Cl_{1,2}$,
\[
[\widehat{\cal W}_\alpha,\widehat{\cal W}_\beta]_+=-\frac{1}{2}
\eta_{\alpha\beta},\quad
[\widehat{\cal N}_\alpha,\widehat{\cal N}_\beta]_+=
-\frac{1}{2}\eta_{\alpha\beta}.
\]
The projector operator $\Pi_-=1-\Pi_+$ commutes with all
the observable operators (\ref{ob0}). 
As a result, one can construct operators 
$\widehat{\cal R}_\alpha=\Pi_-\widehat{\cal W}_\alpha$
and $\widehat{\cal U}_\alpha=\Pi_-\widehat{\cal N}_\alpha$
forming the $su(1,1)$ algebras but
giving the  $s(1,2)$ superalgebras \cite{CrRit}
\[
[\widehat{\cal R}_\alpha,\widehat{\cal R}_\beta]_+=-\frac{1}{2}
\eta_{\alpha\beta}\Pi_-,\quad
[\Pi_-,\widehat{\cal R}_\alpha]=0,
\]
\[
[\widehat{\cal U}_\alpha,
\widehat{\cal U}_\beta]_+=-\frac{1}{2}\eta_{\alpha\beta}\Pi_-,
\quad
[\Pi_-,\widehat{\cal U}_\alpha]=0
\]
instead of Clifford algebras $Cl_{1,2}$.
Operators $\widehat{\cal R}_\alpha$ were revealed in Ref. \cite{gps}
as integrals of motion of $P,T$-invariant planar fermion
system:
\begin{eqnarray}
&\widehat{\cal R}{}^0=\frac{1}{2}(\widehat{S}+\widehat{N})=
-\frac{1}{4}(\gamma^{(0)}\otimes 1
-1\otimes \sigma_3),&\nonumber\\
&
\widehat{\cal R}{}^1=\frac{1}{2}(K^{(1)}-T^{(2)})=
\frac{1}{4}(\gamma^{(2)}\otimes\sigma_1-
\gamma^{(1)}\otimes\sigma_2),
&\nonumber\\
&\widehat{\cal R}{}^2=\frac{1}{2}(K^{(2)}+T^{(1)})=
-\frac{1}{4}(\gamma^{(1)}\otimes\sigma_1+
\gamma^{(2)}\otimes\sigma_2).&
\label{rr}
\end{eqnarray}
Operator $\Pi_-$ turns
into unity on the physical subspace (\ref{3dir}).
As a result, operators 
$\widehat{\cal R}_\alpha$ and 
$\widehat{\cal U}_\alpha$, like $\widehat{\cal W}_\alpha$
and $\widehat{\cal N}_\alpha$,
generate spin-1/2 representation of SU(1,1).

One could conclude that the described symmetries 
(generated by the corresponding observable operators) 
are trivial. Indeed, $\widehat{S}$ is the spin operator
related to $\widehat{N}$ via the basic equation 
(\ref{3dir}) and $\widehat{T}{}^i$, $\widehat{K}{}^i$ simply
mix `up' (spin $s=-1/2$) and `down' 
(spin $s=+1/2$) states. 
However, the non-triviality of these symmetries consists
in the fact that the operators $\widehat{T}{}^i$
and $\widehat{K}{}^i$ are not Lorentz scalars.
Their covariant counterparts
can be presented, e.g., in the vector form
\[
\tilde{\cal T}_\mu=\epsilon_{\mu\nu\lambda}p^\nu\gamma^\lambda
\otimes\sigma_1=
\frac{\sqrt{2}}{m}\epsilon_{\mu\nu\lambda}p^\nu\widehat{\Xi}{}^\lambda,
\quad
\tilde{\cal K}_\mu=\epsilon_{\mu\nu\lambda}p^\nu\gamma^\lambda
\otimes\sigma_2=2i\tilde{\cal T}_\mu\widehat{N},
\] 
$p_\mu\tilde{\cal T}{}^\mu=p_\mu
\tilde{\cal K}{}^\mu=0.
$
These (related by $\widehat{N}$) covariant operators act non-trivially in
the fermion spaces specified by the corresponding
(omitted)
spinor indices of $\psi_u$ and $\psi_d$ and
in `isotopic' space described by $u,d$ subindices,
and being linear in $p_\mu$,
act also on the space-time argument of the state
$\Psi(x)$. Their (anti)commutators,
\[
[\tilde{\cal T}_\mu,\tilde{\cal T}_\nu]=2(p^2\eta_{\mu\nu}-
p_\mu p_\nu),\quad
[\tilde{\cal T}_\mu,\tilde{\cal T}_\nu]_{{}_+}=
4i\epsilon_{\mu\nu\lambda}p^\lambda(p\widehat{J}),
\]
are nonlinear in Poincar\'e generators.

Actually, as was noted in Refs. \cite{ps,gps}, the $P,T$-invariant
fermion system can be considered as the system realizing irreducible
representation of the nonstandard super-extension of the $3D$
Poincar\'e group characterized by zero superspin.  Let us describe
such super-extension associated to the $P,T$-invariant planar fermion
system.  Here we have two related possibilities in correspondence
with non-covariant (super)algebraic relations described above.
First, the generators of $3D$ Poincar\'e group, $p_\mu$ and
$\widehat{J}_\mu$, can be supplemented by the generators $\tilde{\cal
W}_\mu=e^{(\alpha)}_\mu(p)
\widehat{W}_\alpha$ (or by the generators
$\tilde{\cal N}_\mu=e^{(\alpha)}_\mu(p)
\widehat{N}_\alpha$).
Then in addition to the $3D$ Poincar\'e algebra, $[p_\mu,p_\nu]=0$,
$[\widehat{J}_\mu,\widehat{J}_\nu]=-i\epsilon_{\mu\nu\lambda}
\widehat{J}{}^\lambda$, 
$[\widehat{J}_\mu,p_\nu]=-i\epsilon_{\mu\nu\lambda} p^\lambda$,
we have the (anti)commutation relations
\[
[\tilde{\cal W}_\mu,\tilde{\cal W}_\nu]=-
i\epsilon_{\mu\nu\lambda}\tilde{\cal W}{}^\lambda,\quad
[\tilde{\cal W}_\mu,\tilde{\cal W}_\nu]_{{}_+}=
-\frac{1}{2}\eta_{\mu\nu},
\]
\[
[\widehat{J}_\mu,\tilde{\cal W}_\nu]=-i
\epsilon_{\mu\nu\lambda}\tilde{\cal W}{}^\lambda,\quad
[p_\mu, \tilde{\cal W}_\nu]=0,
\]
This gives us the super-extension of
the Poincar\'e algebra characterized by the Casimir operators
$p^2$ and ${\cal S}=e^{(0)}_\mu{\cal J}{}^\mu$,
${\cal J}_\mu=\widehat{J}_\mu-\tilde{\cal W}_\mu$,
$[{\cal J}_\mu,{\cal J}_\nu]=
-i\epsilon_{\mu\nu\lambda}{\cal J}^\lambda$.
The operator ${\cal S}$ has a sense of superspin.  
Under taking into account the concrete realization
of $\tilde{\cal W}_\mu$, we find that it identically 
turns into zero.

Another possibility to construct the 
super-extension of the Poincar\'e algebra
consists in supplementing the set of Poincar\'e generators
by the covariant set of generators
$\Pi_-$ and $\tilde{\cal R}_\mu$ (or by $\tilde{\cal U}_\mu$
instead of $\tilde{\cal R}_\mu$), 
satisfying the superalgebraic relations
\[
[\tilde{\cal R}_\mu,\tilde{\cal R}_\nu]=
-i\epsilon_{\mu\nu\lambda}\tilde{\cal R}{}^\lambda,\quad
[\tilde{\cal R}_\mu,\tilde{\cal R}_\nu]_{{}_+}=
-\frac{1}{2}\Pi_-\eta_{\mu\nu},\quad
[\tilde{\cal R}_\mu,\Pi_-]=0,
\]
\[
[\widehat{J}_\mu,\tilde{\cal R}_\nu]=
-i \epsilon_{\mu\nu\lambda}\tilde{\cal R}{}^\lambda,
\quad
[p_\mu,\tilde{\cal R}_\nu]=[\Pi_-,p_\mu]=
[\Pi_-,\widehat{J}_\mu]=0.
\]
Here $\tilde{\cal R}_\mu=e^{(\alpha)}_\mu(p)\widehat{R}_\alpha$
and $\Pi_-=\frac{1}{2}+2\widehat{S}\widehat{N}$.
In this case the superspin Casimir operator
is given by ${\cal S}=e^{(0)}_\mu{\cal J}{}^\mu$, with
${\cal J}_\mu=\widehat{J}_\mu-\tilde{\cal R}_\mu$.
The superspin takes here zero value on the physical subspace 
specified by Eq. (\ref{3dir}).
This second form of super-extension of the Poincar\'e algebra was
discussed in Refs. \cite{ps,gps} as a hidden supersymmetry
of the $P,T$-invariant planar fermion system.

Having in mind that in $2+1$ dimensions
the transformations of space inversion
are realized as \cite{gps}
$P:\Psi(x)\rightarrow
\Psi'(x')=\eta U_P\Psi(x),$
$U_P=\gamma^1\otimes\sigma_1$,
$x'_\mu=(x_0,-x_1,x_2)$, $\vert\eta\vert=1$,
we find that the operator $\widehat{N}$ is parity-odd,
$[\widehat{N},U_P]_{{}_+}=0$, and, being a Lorentz scalar,
has a sense of changing parity operator.  The given quantum
$P$-transformations correspond to the classical transformations of
space inversion accepted in BM, BCL and CPV models.

\section{Classical symmetries and anomalies}
Let us find the classical even and odd integrals of motion
corresponding to quantum observables.  For BCL model the
correspondence is the following:
\[
\widehat{S}\rightarrow S_e=-\frac{i}{2}
\epsilon_{ij}\xi^{(i)}\xi^{(j)},\quad
S_o=-\sqrt{2}\theta_0\theta_1\xi^{(0)},
\]
\[
\widehat{N}\rightarrow
N_e=-\frac{i}{3}
\theta_1\epsilon_{\mu\nu\lambda}\xi^\mu\xi^\nu\xi^\lambda,\quad
N_o=\frac{1}{\sqrt{2}}
\theta_0,
\]
\[
\widehat{T}{}^i\rightarrow
T_e^{i}=-2i\theta_0\theta_1\xi^{(0)}\epsilon^{ij}\xi^{(j)},\quad
T_o^{i}=-\frac{1}{\sqrt{2}}
\xi^{(i)},
\]
\[
\widehat{K}{}^{i}\rightarrow
K^{i}_e=i\theta_0\xi^{(i)},\quad
K^{i}_o=-\sqrt{2}\theta_1\xi^{(0)}\epsilon^{ij}\xi^{(j)}.
\]
One can check that $S_e$ and $K^{i}_e$
reproduce $su(1,1)$ algebra with respect 
to classical brackets.
The classical integrals
$N_o$ and $T^{i}_o$ reproduce
classically the analog of $Cl_{1,2}$ algebra.
So, classically, unlike the quantum case,
$su(1,1)$ and $Cl_{1,2}$ in BCL model are generated by different sets
of integrals of motion.
On the other hand, though $S_e$ and $T^{i}_e$
commute with respect to the brackets
in appropriate way, the quantities $T^{i}_e$, $i=1,2$, 
satisfy the relation $\{T^{1}_e,T^{2}_e\}=0$,
and therefore, this set does not reproduce classically $su(1,1)$.
Note also that  $S_o$, $N_e$, $T^i_e$ and
$K^i_o$ are pure imaginary  classical  quantities.
Thus, we can reproduce here classically
$su(1,1)$ and $Cl_{1,2}$, but, nevertheless, we cannot reproduce all
the (anti)commutation relations of the corresponding quantum integrals.
The asymmetry between classical analogs of operators $\widehat{T}{}^i$
and $\widehat{K}{}^i$ has a simple reason: though the basic equation
(\ref{3dir}) is `symmetric' in indices $a=1,2$ of $\sigma$-matrices,
the concrete realization of odd operators (\ref{?7}) in the model
destroys this symmetry.

Let us multiply classical generators of $su(1,1)$ and $Cl_{1,2}$  by
$C=\frac{1}{2}+\xi^{(0)}\theta_1$, which commutes with all the
classical integrals of motion listed above, i.e. is a central element
weakly equal to $\frac{1}{2}$.  As a result we get classically
$s(1,2)$ superalgebra instead of Clifford algebra $Cl_{1,2}$, but
destroy $su(1,1)$ bracket relations of even generators since $C$
being the classical analog of the operator $\Pi_-$ does not reproduce
its defining property $\Pi_-^2=\Pi_-$.  Therefore, though in BCL
model we can reproduce classically $s(1,2)$ superalgebra, its
generators, unlike the quantum generators of $s(1,2)$ superalgebra,
are not related at all to $su(1,1)$ generators.

As we remember, the BM model is different from the BCL model because
of absence of classical odd integral $\theta_0$.  Since the sets of
even, $S_e$, $K^{i}_e$, and  odd, $N_o$, $T^{i}_o$, integrals of
motion of BCL model contain $\theta_0$, classically either $su(1,1)$
or $Cl_{1,2}$ cannot be reproduced in BM model.  So, in BM model even
a part of quantum symmetries cannot be reproduced classically.
 
Let us turn to the CPV model.
Here formally the direct classical analogs of all the
listed above quantum observables can be constructed:
\[
\widehat{S}\rightarrow S_e=-\frac{i}{2}
\epsilon_{ij}\xi^{(i)}\xi^{(j)},\quad
S_o=i\sqrt{2}\theta_1\theta_2\xi^{(0)},
\]
\[
\widehat{N}\rightarrow
N_e=-i\theta_1\theta_2,\quad
N_o=-\frac{i}{3\sqrt{2}}
\epsilon_{\mu\nu\lambda}\xi^{\mu}\xi^\nu\xi^\lambda,
\]
\[
\widehat{T}{}^{i}\rightarrow
T_e^{i}=i\theta_2\xi^{(i)},\quad
T_o^{i}=-i\sqrt{2}\theta_1\xi^{(0)}\epsilon^{ij}\xi^{(j)},
\]
\[
\widehat{K}{}^{i}\rightarrow
K^{i}_e=-i\theta_1\xi^{(i)},\quad
K^{i}_o=-i\sqrt{2}
\theta_2\xi^{(0)}\epsilon^{ij}\xi^{(j)}.
\]
All these quantities are real,
but not all of them are classical integrals of motion:
$T^{i}_e$, $T^{i}_o$, $K^{i}_e$ and $K^{i}_o$
are not conserved.
Only their specific linear combinations,
$K^{1}_e-T^{2}_e$, $K^{2}_e+T^{1}_e$,
and $K^{1}_o-T^{2}_o$, $K^{2}_o+T^{1}_o$,
are integrals of motion.
One can check that even linear combinations
$\frac{1}{2}(K^{1}_e-T^{2}_e)$, 
$\frac{1}{2}(K^{2}_e+T^{1}_e)$,
together with even linear combination 
$\frac{1}{2}(S_e+N_e)$, being classical analogs of quantum 
linear combinations (\ref{rr}), reproduce classically
$su(1,1)$.
On the other hand, the odd linear combinations
$\frac{1}{2}(K^{1}_o-T^{2}_o)$ and 
$\frac{1}{2}(K^{2}_o+T^{1}_o)$,
together with odd linear combination 
$\frac{1}{2}(S_o+N_o)$
reproduce classical analog of $s(1,2)$ superalgebra.
Note that these classical generators of $s(1,2)$ 
are simply $su(1,1)$ generators multiplied by the odd
integral $-\sqrt{2}\xi^{(0)}$. At the quantum level
the same operators correspond to even and odd
above-mentioned integrals of motion, and, as a consequence,
operators $\widehat{\cal R}_\alpha$ satisfy simultaneously
$su(1,1)$ commutation and $s(1,2)$ anticommutation relations.

Therefore, we see that the CPV model reproduces classically
$su(1,1)$ and $s(1,2)$, but fails to reproduce $Cl_{1,2}$. 
What is a reason of such difference between 
quantum and classical case in this model?
In quantum case the observables $\widehat{T}{}^{i}$
and $\widehat{K}{}^{i}$ anticommute with the 
quantum operator corresponding to the 
classical constraint $\varrho$.
Therefore, from this point of view, they form
some sort of superalgebra with $\widehat{\varrho}$.
But classically $\varrho$ is the even quantity.
Due to this, the quantum anticommutation relations of
$\widehat{T}{}^{i}$ and $\widehat{K}{}^{i}$ with $\widehat{\varrho}$
cannot be
reproduced classically with respect to the bracket. This results in
the fact that only part of quantum observables has classical analogs
and, as a consequence, not all the quantum symmetry superalgebraic
relations can be reproduced classically in CPV model.

\section{U${}_{\sigma_3}$(1) gauge symmetry}
As we noted,
the CPV model is invariant under global $U(1)$ transformations
(\ref{u1})
generated by the integral $N_\theta=-i\theta_1\theta_2$.
Gauging this symmetry, $\omega\rightarrow
\omega(x_\mu(\tau))$, we arrive at the 
Lagrangian \cite{gps}
\begin{equation}
L^g_{CPV}=L_{CPV}+qN_\theta\left(\dot{x}_\mu{\cal A}^\mu+\frac{e}{2}
i\xi_\mu\xi_\nu{\cal F}^{\mu\nu}\right)
\label{lcpvg}
\end{equation}
being invariant with respect to the local U(1) transformations:
\[
\theta^\pm
\rightarrow\theta^{\pm}{}'=e^{\pm i\omega(x)}\theta^\pm,\quad
{\cal A}_\mu\rightarrow
{\cal A}'_\mu={\cal A}_\mu-q^{-1}\partial_\mu\omega(x).
\]
Here ${\cal A}_\mu={\cal A}_\mu(x)$ is a U(1) gauge field,
$q$ is a coupling constant and ${\cal F}_{\mu\nu}=
\partial_\mu{\cal A}_\nu-\partial_\nu{\cal A}_\mu$.
Lagrangian (\ref{lcpvg}) leads to the modified
(gauged) constraints
\begin{equation}
\phi^g=\frac{1}{2}({\cal P}^2+m^2-iqN_\theta\xi_\mu\xi_\nu
{\cal F}^{\mu\nu})\approx 0,\quad
\varrho^g=i\epsilon_{\mu\nu\lambda}{\cal P}^\mu\xi^\nu\xi^\lambda
-2mN_\theta\approx 0,
\label{cong}
\end{equation}
where ${\cal P}_\mu=p_\mu-qN_\theta{\cal A}_\mu$.
These constraints form the same trivial algebra,
$\{\phi^g,\varrho^g\}=0$, which takes  place in a free
case.
The quantum counterparts of classical constraints
(\ref{cong}) have the form 
\begin{equation}
\widehat{\varrho}{}^g\Psi=0,\quad
\widehat{\phi}{}^g\Psi=0,
\label{eg}
\end{equation}
\begin{equation}
\widehat{\varrho}{}^g=\widehat{\cal P}\gamma\otimes 1+
m\cdot1\otimes\sigma_3,\quad
2\widehat{\phi}{}^g=\widehat{\cal P}{}^2+m^2+\frac{1}{4}q
\epsilon_{\mu\nu\lambda}{\cal F}^{\mu\nu}\gamma^\lambda\otimes 1,
\label{gco}
\end{equation}
where $\widehat{\cal P}_\mu=p_\mu-\frac{1}{2}q{\cal A}_\mu(x)
1\otimes\sigma_3$.
The quantum constraints reproduce the trivial 
algebra, $[\widehat{\phi}{}^g,\widehat{\varrho}{}^g]=0$,
and as in a free case, satisfy the relation
$\widehat{\varrho}{}^g{}^2=-2\widehat{\phi}{}^g+2m(1\otimes\sigma_3)
\widehat{\varrho}{}^g$ which says that the gauged mass shell condition is
a consequence of the gauged constraint $\widehat{\varrho}{}^g$.  The
explicit form of the quantum constraints (\ref{gco}) means that the
U${}_{\sigma_3}$(1) gauge interaction is the usual U(1) gauge
interaction with the only difference that spin $s=-1/2$ and $s=+1/2$
states have the coupling constants of the opposite sign.  This
specific form of U(1) gauge interaction was used for
modelling high-temperature superconductivity \cite{highT}.  Thus, we
conclude that the localization of the global U(1) symmetry of CPV
model gives in a natural way the U${}_{\sigma_3}$(1) gauge theory for
$P,T$-invariant planar fermion system.

There is no natural analog of global U(1) symmetry (\ref{u1}) in BM
and BCL models. To reproduce classically U${}_{\sigma_3}$(1)
gauge symmetry in these models, we construct direct classical analogs
of quantum constraint operators (\ref{gco}).  The necessary
modification of odd constraint in BM model is
\begin{equation}
\chi^g=p\xi+\frac{i}{2}q\epsilon_{\mu\nu\lambda}
{\cal A}^\mu\xi^\nu\xi^\lambda\xi_*-m\xi_*\approx 0.
\label{tich}
\end{equation}
In BCL model the modified constraint has the same form
with the change $\xi_*$ for $\theta_1$.
The quantum analog of constraint (\ref{tich})
is exactly the first quantum equation from Eqs. (\ref{eg})
(multiplied by $\frac{1}{\sqrt{2}}1\otimes \sigma_1$).
To find the modified even constraint,
we calculate the bracket of odd constraint 
with itself,
$
\{\chi^g,\chi^g\}=2i\tilde{\phi},
$ 
and get
\begin{equation}
2\tilde{\phi}=p^2+(m-\frac{i}{2}q\epsilon_{\mu\nu\lambda}
{\cal A}^\mu\xi^\nu\xi^\lambda)^2-
\frac{1}{3}q\epsilon_{\mu\nu\lambda}\xi^\mu\xi^\nu\xi^\lambda
\xi_*\partial_\sigma{\cal A}^\sigma
-iq\epsilon_{\mu\nu\lambda}(p^\mu{\cal A}^\nu+
{\cal A}^\nu p^\mu)\xi^\lambda\xi_*\approx 0.
\label{tilf}
\end{equation}
Due to Jacobi identity we have $\{\chi^g,\tilde{\phi}\}=0$ and
conclude that the modified constraints form the same $s(1)$
superalgebra as in a free case.  Note that in BM and BCL models,
unlike the CPV model, classical even constraint contains no term
proportional to ${\cal F}_{\mu\nu}$.  This term appears only quantum
mechanically, under construction of the quantum analog of the
constraint (\ref{tilf}).  Indeed, choosing the same ordering for
quantum counterparts of the classical quantities presenting in Eq.
(\ref{tilf}), we get
\[
2\widehat{\tilde{\phi}}=2\widehat{\phi}{}^g-q{\cal A}\gamma\otimes\sigma_3
\cdot\widehat{\varrho}{}^g,
\]
where $\widehat{\phi}{}^g$ and $\widehat{\varrho}{}^g$ are given by Eq.
(\ref{gco}).  Let us stress that to get the correct form of the
quantum constraint $\widehat{\tilde{\phi}}$, it is essential to take into
account the term $-\frac{1}{4}q^2(\epsilon_{\mu\nu\lambda} {\cal
A}^\mu\xi^\nu\xi^\lambda)^2$ which appears from the second term in
Eq. (\ref{tilf}) and classically is equal to zero.
Moreover, for getting the appropriate quantum analog of even
constraint, it is essential to preserve in (\ref{tilf}) the term
$\frac{1}{3}q\epsilon_{\mu\nu\lambda}\xi^\mu\xi^\nu\xi^\lambda
\xi_*\partial_\sigma{\cal A}^\sigma$
in spite of the fact that classically it is 
proportional to $\chi^g$, and so, itself is weakly equal
to zero. 

{}From here we conclude that the quantization
of even constraint gives the necessary gauged 
quantum mass shell condition, but
the relationship of classical and
quantum theories for the U${}_{\sigma_3}$(1) gauged BM and BCL
models is not direct and natural.

Concluding the discussion of classical
U${}_{\sigma_3}$(1) gauge symmetry, let us write explicitly
the corresponding modified Lagrangians for BM and BCL models
which can be obtained by inverse Legendre transformation
proceeding from the classical constraints (\ref{tich}) and (\ref{tilf}):
\begin{equation}
L^g_{BM}=\frac{1}{2e}(\dot{x}_\mu-i\lambda\xi_\mu)^2
-\frac{e}{2}m_-^2 +i\lambda m_+\xi_*
+iq\epsilon_{\mu\nu\lambda}\dot{x}{}^\mu{\cal A}^\nu\xi^\lambda
\xi_*
-\frac{i}{2}\xi\dot{\xi}-\frac{i}{2}\xi_*\dot{\xi}_*,
\label{lbmg}
\end{equation}
\begin{equation}
L^g_{BCL}=\frac{1}{2e}\dot{x}{}^2-\frac{e}{2}m_-^2
+\frac{i}{em_+}(\dot{\xi}_*+q\epsilon_{\mu\nu\lambda}
\dot{x}{}^{\mu}{\cal A}^{\nu}\xi^\lambda)\dot{x}\xi
-\frac{i}{2}\xi_\mu\dot{\xi}{}^\mu+
\frac{i}{2}\xi_*\dot{\xi}_*,
\label{lbclg}
\end{equation}
where $m_\pm=m\pm\frac{i}{2}q\epsilon_{\mu\nu\lambda}
{\cal A}^\mu\xi^\nu\xi^\lambda$.
The comparison of the Lagrangians
(\ref{lcpvg}), (\ref{lbmg}) and (\ref{lbclg})
confirms the simplicity and naturalness of classical analog 
of U${}_{\sigma_3}$(1) gauge theory for CPV model 
but its rather obscure and unnatural 
character for the case of BM and BCL models.

\section{Outlook}

We have observed the phenomenon of classical anomaly for three
pseudoclassical models of $3D$ $P,T$-invariant system of planar
fermions.  One of the models is the CPV model having nilpotent
constraint being nonlinear in Grassmann variables and admitting
no, even local, gauge conditions.  Because of this, the CPV model
admits the quantization only by the Dirac method whereas the reduced
phase space quantization cannot be applied to it.  The application of
path-integral method seems also to be problematic for such
class of constrained systems \cite{pr}. On the other hand, BM
and BCL models contain linear in Grassmann variables nilpotent
constraint and path-integral quantization method can in
principle (see below) be applied for them.  There are other
pseudoclassical models containing nilpotent nonlinear
constraints \cite{pseudo1,pseudo2}.  Such models after
quantization by the Dirac method describe the vector and higher
spin fields. At least for one of such models, the model of
$P,T$-invariant system of topologically massive vector U(1)
gauge fields \cite{pseudo2}, some elements of classical anomaly
were observed in Ref. \cite{pseudo3}.  Therefore, the natural
question is  whether it is possible to construct for any of
peculiar higher spin  models
\cite{pseudo1,pseudo2} the `supplementary' pseudoclassical model
which would contain only linear nilpotent constraints but under
quantization by the Dirac method would give the same quantum system
as the corresponding known peculiar pseudoclassical model.  If this
question can be answered positively, the phenomenon of classical
anomaly will be revealed for corresponding (equivalent quantum
mechanically) pseudoclassical models.

In the same context it would be interesting to investigate different
classical field theoretical models with odd fermions and also 
answer the intriguing question whether the phenomenon of classical
anomaly can be observed for the models containing no Grassmann
variables.
 
In conclusion we note that since the path-integral quantization
method is closely related to the classical theory of the
corresponding system to be quantized, it seems that the
classical anomaly should reveal itself in some nontrivial way in
path-integral approach to quantum mechanics.  So, it seems to be
very interesting to apply this method for quantizing the $3D$ BM
and BCL models.
\vskip 0.3cm
{\bf Acknowledgements}

J. G. was partially supported by grants 1950278 and 1960229 by
FONDECYT-Chile and by grant 04-953/GR from Dicyt-USACH. M.P. thanks Prof.
C. Teitelboim for discussions and University of Santiago of Chile, where
the part of this work has been realized, for hospitality. One of us (J.G.)
is a recipient of a John S. Guggenheim Fellowship.

\appendix

\section{Appendix}

Here we discuss the quantization of the BM and BCL models in 
$D=2n$ case.
In the BM model, the quantum counterparts of odd variables 
can be realized as
\begin{equation}
\widehat{\xi}_\mu=\frac{1}{\sqrt{2}}\gamma_*\gamma_\mu,\quad
\widehat{\xi}_*=\frac{1}{\sqrt{2}}\gamma_*,
\label{?6}
\end{equation}
where $\gamma_\mu$ are Dirac $2^n\times 2^n$-matrices
satisfying the relations 
$[\gamma_\mu,\gamma_\nu]_{+}=-2\eta_{\mu\nu}$,
$\gamma_\mu^\dagger=-\eta_{\mu\mu}\gamma_\mu$
(no summation),
whereas
$\gamma_*=i^{n}\gamma_0\gamma_1...\gamma_{d-1}$,
$\gamma_*^2=-1$,
$\gamma_*^\dagger=-\gamma_*$.
As a result, the quantum analog
of odd constraint coincides with the 
$2n$-dimensional Dirac equation
multiplied by nonsingular factor $\frac{1}{\sqrt{2}}\gamma_*$,
and we arrive at the conclusion that 
the quantization of BM model in the case of $2n$-dimensional
space-time gives us the corresponding Dirac equation.
The indefinite scalar product $\langle\Psi_1\vert\Psi_2\rangle=
\bar{\Psi}_1\Psi_2$
with $\bar{\Psi}=\Psi^\dagger\Delta$,
$\Delta=\sqrt{2}\widehat{\xi}{}^0$, guarantees that
all the operators $\widehat{\xi}_\mu$, $\widehat{\xi}_*$, and as a 
consequence, all the observables are self-conjugate 
operators. The averaged odd constraint,
$\langle\Psi\vert\widehat{\chi}\Psi\rangle=\Psi^\dagger\gamma^0
(p\gamma+m)\Psi$, gives
us the usual Lagrangian for $2n$-dimensional massive Dirac fermion
field.

The BCL model in $2n$-dimensional space-time has $2n+2$ odd
phase space variables, i.e.  one more in comparison with BM
model.  This difference is crucial for realization of quantum
analogs of odd variables.  The dimension of irreducible
representation of the Clifford algebra with $2n+2$ generators is
$2^{n+1}$, twice more of that we had in BM model.  As a
result, quantum mechanically BM and BCL models are not
equivalent in the case of $D=2n$. Here
one can realize odd operators in two different 
forms related by the corresponding unitary transformation.
First, they can be realized as
\begin{equation}
\widehat{\xi}_\mu=\frac{1}{\sqrt{2}}\gamma_*\gamma_\mu\otimes
\sigma_1,\quad
\widehat{\theta}_1=\frac{1}{\sqrt{2}}\gamma_*\otimes\sigma_1,\quad
\widehat{\theta}_0=\frac{1}{\sqrt{2}}1\otimes\sigma_3.
\label{real1}
\end{equation}
In this case the quantum analog of the odd constraint gives 
two $2n$-dimensional Dirac equations (multiplied 
by nonsingular factor $\frac{1}{\sqrt{2}}\gamma_*\otimes\sigma_1$),
and corresponding fermionic states are distinguished
by the operator $\widehat{\theta}_0$.
On the other hand, if we realize the Clifford algebra generators
as
\begin{equation}
\widehat{\xi}_\mu=\frac{1}{\sqrt{2}}\gamma_\mu\otimes\sigma_1,\quad
\widehat{\theta}_1=\frac{i}{\sqrt{2}}1\otimes\sigma_2,\quad
\widehat{\theta}_0=\frac{1}{\sqrt{2}}1\otimes\sigma_3,
\label{real2}
\end{equation}
the quantum counterpart of the odd constraint equation
gives us the equation of the form (\ref{D3}).
Realizations (\ref{real1}) and (\ref{real2}) are related by the
unitary transformation $S\widehat{G}S^{-1}=\widehat{G}{}'$,
where $S=i(\gamma_*\otimes P_+-1\otimes P_-)$,
$P_\pm=\frac{1}{2}(1\pm\sigma_3)$,
and $\widehat{G}$ and $\widehat{G}{}'$ are corresponding operators
in realization (\ref{real1}) and (\ref{real2}), respectively.

Choosing the scalar product with indefinite
metric operator $\Delta=\sqrt{2}\widehat{\xi}{}^0$, we get the operators
$\widehat{\xi}_\mu$ and $\widehat{\theta}_1$ as self-conjugate ones.
As a consequence, observables $\widehat{J}_{\mu\nu}$ and
$\widehat{\Xi}_\mu$ will also be self-conjugate together with the
quantum odd constraint which  under averaging supplies us with
the corresponding form of the Lagrangian, ${\cal
L}=\Psi^\dagger\gamma^0\otimes1(p\gamma\otimes1 +m\cdot 1\otimes
\Gamma)\Psi$, where $\Gamma=1$ and $\Gamma=\sigma_3$ for
realizations (\ref{real1}) and (\ref{real2}).  But as in the
odd-dimensional case, the model again reveals the quantum
anomaly: the quantum counterpart of the classical real
observable variable $\theta_0$ turns out to be anti-self-conjugate. 

Concluding, we note that in the case of $(3+1)$-dimensional BM
model, three  nontrivial transverse components
$\widehat{\Xi}{}^\perp_\mu$,
$\widehat{\Xi}{}^\perp_\mu p^\mu=0$, of the observable vector
$\widehat{\Xi}_\mu$ are, in fact, the components of the
Pauli-Lubanski vector for the Dirac field.  On the other hand,
the quantum BCL model gives us the pair of Dirac fields.  In
representation (\ref{real1}) the matrix part of the parity
operator is $U_P=\gamma^0\otimes\sigma_1$.  Therefore, the
additional operator $\widehat{\theta}_0$ has here, as in the
case of $2+1$ dimensions, the sense of parity-changing operator.
As a result, the quantum BCL model, unlike the BM model,
describes the pair of massive Dirac fermion fields having
opposite internal parities.

\end{document}